\documentclass[twocolumn,letterpaper,aps,prc,superscriptaddress,showpacs,nofootinbib,floatfix]{revtex4}

\usepackage{graphicx} 	
\usepackage{dcolumn} 	
\usepackage{bm} 	

\begin{document}

\title{Production of $\omega$ mesons in $p+p$, $d+$Au, Cu$+$Cu, and Au$+$Au collisions \\
at $\sqrt{s_{NN}}$=200 GeV }

\newcommand{\abilene}{Abilene Christian University, Abilene, Texas 79699, USA}
\newcommand{\banaras}{Department of Physics, Banaras Hindu University, Varanasi 
221005, India}
\newcommand{\barc}{Bhabha Atomic Research Centre, Bombay 400 085, India}
\newcommand{\bnlcoll}{Collider-Accelerator Department, Brookhaven National 
Laboratory, Upton, New York 11973-5000, USA}
\newcommand{\bnlphys}{Physics Department, Brookhaven National Laboratory, Upton, 
New York 11973-5000, USA}
\newcommand{\caucr}{University of California - Riverside, Riverside, California 
92521, USA}
\newcommand{\charlesczech}{Charles University, Ovocn\'{y} trh 5, Praha 1, 116 36, 
Prague, Czech Republic}
\newcommand{\chonbuk}{Chonbuk National University, Jeonju, 561-756, Korea}
\newcommand{\ciae}{China Institute of Atomic Energy (CIAE), Beijing, People's 
Republic of China}
\newcommand{\cns}{Center for Nuclear Study, Graduate School of Science, University 
of Tokyo, 7-3-1 Hongo, Bunkyo, Tokyo 113-0033, Japan}
\newcommand{\colorado}{University of Colorado, Boulder, Colorado 80309, USA}
\newcommand{\columbia}{Columbia University, New York, New York 10027 and Nevis 
Laboratories, Irvington, New York 10533, USA}
\newcommand{\czechtech}{Czech Technical University, Zikova 4, 166 36 Prague 6, 
Czech Republic}
\newcommand{\dapnia}{Dapnia, CEA Saclay, F-91191, Gif-sur-Yvette, France}
\newcommand{\debrecen}{Debrecen University, H-4010 Debrecen, Egyetem t{\'e}r 1, 
Hungary}
\newcommand{\elte}{ELTE, E{\"o}tv{\"o}s Lor{\'a}nd University, H - 1117 Budapest, 
P{\'a}zm{\'a}ny P. s. 1/A, Hungary}
\newcommand{\ewha}{Ewha Womans University, Seoul 120-750, Korea}
\newcommand{\fit}{Florida Institute of Technology, Melbourne, Florida 32901, USA}
\newcommand{\fsu}{Florida State University, Tallahassee, Florida 32306, USA}
\newcommand{\gsu}{Georgia State University, Atlanta, Georgia 30303, USA}
\newcommand{\hiroshima}{Hiroshima University, Kagamiyama, Higashi-Hiroshima 
739-8526, Japan}
\newcommand{\ihepprot}{IHEP Protvino, State Research Center of Russian 
Federation, Institute for High Energy Physics, Protvino, 142281, Russia}
\newcommand{\illuiuc}{University of Illinois at Urbana-Champaign, Urbana, 
Illinois 61801, USA}
\newcommand{\inrras}{Institute for Nuclear Research of the Russian Academy of 
Sciences, prospekt 60-letiya Oktyabrya 7a, Moscow 117312, Russia}
\newcommand{\instpasczech}{Institute of Physics, Academy of Sciences of the Czech 
Republic, Na Slovance 2, 182 21 Prague 8, Czech Republic}
\newcommand{\isu}{Iowa State University, Ames, Iowa 50011, USA}
\newcommand{\jinrdubna}{Joint Institute for Nuclear Research, 141980 Dubna, 
Moscow Region, Russia}
\newcommand{\jyvaskyla}{Helsinki Institute of Physics and University of 
Jyv{\"a}skyl{\"a}, P.O.Box 35, FI-40014 Jyv{\"a}skyl{\"a}, Finland}
\newcommand{\kaeri}{KAERI, Cyclotron Application Laboratory, Seoul, Korea}
\newcommand{\kek}{KEK, High Energy Accelerator Research Organization, Tsukuba, 
Ibaraki 305-0801, Japan}
\newcommand{\kfki}{KFKI Research Institute for Particle and Nuclear Physics of the 
Hungarian Academy of Sciences (MTA KFKI RMKI), H-1525 Budapest 114, POBox 49, 
Budapest, Hungary}
\newcommand{\korea}{Korea University, Seoul, 136-701, Korea}
\newcommand{\kurchatov}{Russian Research Center ``Kurchatov Institute", Moscow, 
123098 Russia}
\newcommand{\kyoto}{Kyoto University, Kyoto 606-8502, Japan}
\newcommand{\labllr}{Laboratoire Leprince-Ringuet, Ecole Polytechnique, 
CNRS-IN2P3, Route de Saclay, F-91128, Palaiseau, France}
\newcommand{\lawllnl}{Lawrence Livermore National Laboratory, Livermore, 
California 94550, USA}
\newcommand{\losalamos}{Los Alamos National Laboratory, Los Alamos, New Mexico 
87545, USA}
\newcommand{\lpc}{LPC, Universit{\'e} Blaise Pascal, CNRS-IN2P3, Clermont-Fd, 
63177 Aubiere Cedex, France}
\newcommand{\lund}{Department of Physics, Lund University, Box 118, SE-221 00 
Lund, Sweden}
\newcommand{\maryland}{University of Maryland, College Park, Maryland 20742, USA}
\newcommand{\mass}{Department of Physics, University of Massachusetts, Amherst, 
Massachusetts 01003-9337, USA }
\newcommand{\muenster}{Institut fur Kernphysik, University of Muenster, D-48149 
Muenster, Germany}
\newcommand{\muhlenberg}{Muhlenberg College, Allentown, Pennsylvania 18104-5586, 
USA}
\newcommand{\myongji}{Myongji University, Yongin, Kyonggido 449-728, Korea}
\newcommand{\nagasaki}{Nagasaki Institute of Applied Science, Nagasaki-shi, 
Nagasaki 851-0193, Japan}
\newcommand{\newmex}{University of New Mexico, Albuquerque, New Mexico 87131, USA 
}
\newcommand{\nmsu}{New Mexico State University, Las Cruces, New Mexico 88003, USA}
\newcommand{\ornl}{Oak Ridge National Laboratory, Oak Ridge, Tennessee 37831, USA}
\newcommand{\orsay}{IPN-Orsay, Universite Paris Sud, CNRS-IN2P3, BP1, F-91406, 
Orsay, France}
\newcommand{\peking}{Peking University, Beijing, People's Republic of China}
\newcommand{\pnpi}{PNPI, Petersburg Nuclear Physics Institute, Gatchina, 
Leningrad region, 188300, Russia}
\newcommand{\riken}{RIKEN Nishina Center for Accelerator-Based Science, Wako, 
Saitama 351-0198, Japan}
\newcommand{\rikjrbrc}{RIKEN BNL Research Center, Brookhaven National Laboratory, 
Upton, New York 11973-5000, USA}
\newcommand{\rikkyo}{Physics Department, Rikkyo University, 3-34-1 
Nishi-Ikebukuro, Toshima, Tokyo 171-8501, Japan}
\newcommand{\saispbstu}{Saint Petersburg State Polytechnic University, St. 
Petersburg, 195251 Russia}
\newcommand{\saopaulo}{Universidade de S{\~a}o Paulo, Instituto de F\'{\i}sica, 
Caixa Postal 66318, S{\~a}o Paulo CEP05315-970, Brazil}
\newcommand{\seoulnat}{Seoul National University, Seoul, Korea}
\newcommand{\stonybrkc}{Chemistry Department, Stony Brook University, SUNY, Stony 
Brook, New York 11794-3400, USA}
\newcommand{\stonycrkp}{Department of Physics and Astronomy, Stony Brook 
University, SUNY, Stony Brook, New York 11794-3400, USA}
\newcommand{\subatech}{SUBATECH (Ecole des Mines de Nantes, CNRS-IN2P3, 
Universit{\'e} de Nantes) BP 20722 - 44307, Nantes, France}
\newcommand{\tenn}{University of Tennessee, Knoxville, Tennessee 37996, USA}
\newcommand{\titech}{Department of Physics, Tokyo Institute of Technology, 
Oh-okayama, Meguro, Tokyo 152-8551, Japan}
\newcommand{\tsukuba}{Institute of Physics, University of Tsukuba, Tsukuba, 
Ibaraki 305, Japan}
\newcommand{\vandy}{Vanderbilt University, Nashville, Tennessee 37235, USA}
\newcommand{\waseda}{Waseda University, Advanced Research Institute for Science 
and Engineering, 17 Kikui-cho, Shinjuku-ku, Tokyo 162-0044, Japan}
\newcommand{\weizmann}{Weizmann Institute, Rehovot 76100, Israel}
\newcommand{\yonsei}{Yonsei University, IPAP, Seoul 120-749, Korea}
\affiliation{\abilene}
\affiliation{\banaras}
\affiliation{\barc}
\affiliation{\bnlcoll}
\affiliation{\bnlphys}
\affiliation{\caucr}
\affiliation{\charlesczech}
\affiliation{\chonbuk}
\affiliation{\ciae}
\affiliation{\cns}
\affiliation{\colorado}
\affiliation{\columbia}
\affiliation{\czechtech}
\affiliation{\dapnia}
\affiliation{\debrecen}
\affiliation{\elte}
\affiliation{\ewha}
\affiliation{\fit}
\affiliation{\fsu}
\affiliation{\gsu}
\affiliation{\hiroshima}
\affiliation{\ihepprot}
\affiliation{\illuiuc}
\affiliation{\inrras}
\affiliation{\instpasczech}
\affiliation{\isu}
\affiliation{\jinrdubna}
\affiliation{\jyvaskyla}
\affiliation{\kaeri}
\affiliation{\kek}
\affiliation{\kfki}
\affiliation{\korea}
\affiliation{\kurchatov}
\affiliation{\kyoto}
\affiliation{\labllr}
\affiliation{\lawllnl}
\affiliation{\losalamos}
\affiliation{\lpc}
\affiliation{\lund}
\affiliation{\maryland}
\affiliation{\mass}
\affiliation{\muenster}
\affiliation{\muhlenberg}
\affiliation{\myongji}
\affiliation{\nagasaki}
\affiliation{\newmex}
\affiliation{\nmsu}
\affiliation{\ornl}
\affiliation{\orsay}
\affiliation{\peking}
\affiliation{\pnpi}
\affiliation{\riken}
\affiliation{\rikjrbrc}
\affiliation{\rikkyo}
\affiliation{\saispbstu}
\affiliation{\saopaulo}
\affiliation{\seoulnat}
\affiliation{\stonybrkc}
\affiliation{\stonycrkp}
\affiliation{\subatech}
\affiliation{\tenn}
\affiliation{\titech}
\affiliation{\tsukuba}
\affiliation{\vandy}
\affiliation{\waseda}
\affiliation{\weizmann}
\affiliation{\yonsei}
\author{A.~Adare} \affiliation{\colorado}
\author{S.~Afanasiev} \affiliation{\jinrdubna}
\author{C.~Aidala} \affiliation{\columbia} \affiliation{\mass}
\author{N.N.~Ajitanand} \affiliation{\stonybrkc}
\author{Y.~Akiba} \affiliation{\riken} \affiliation{\rikjrbrc}
\author{H.~Al-Bataineh} \affiliation{\nmsu}
\author{A.~Al-Jamel} \affiliation{\nmsu}
\author{J.~Alexander} \affiliation{\stonybrkc}
\author{A.~Angerami} \affiliation{\columbia}
\author{K.~Aoki} \affiliation{\kyoto} \affiliation{\riken}
% ...
\collaboration{PHENIX Collaboration} \noaffiliation

\date{\today}

\begin{abstract}

The PHENIX experiment at the Relativistic Heavy Ion Collider has 
measured $\omega$ meson production via leptonic and hadronic decay 
channels in $p+p$, $d+$Au, Cu$+$Cu, and Au$+$Au collisions at 
$\sqrt{s_{NN}}$=200 GeV.  The invariant transverse momentum spectra 
measured in different decay modes give consistent results. Measurements in 
the hadronic decay channel in Cu$+$Cu and Au$+$Au collisions show that 
$\omega$ production has a suppression pattern at high transverse momentum, 
similar to that of $\pi^{0}$ and $\eta$ in central collisions, but no 
suppression is observed in peripheral collisions.  The nuclear 
modification factors, $R_{\rm AA}$, are consistent in Cu$+$Cu and Au$+$Au 
collisions at similar numbers of participant nucleons.

\end{abstract}

\pacs{25.75.Dw}

\maketitle

\section{\label{sec:int}Introduction}

The measurement of hadrons produced in relativistic heavy-ion collisions 
is a well established tool in the study of the hot and dense matter 
created in the collisions. The PHENIX experiment at the 
Relativistic Heavy Ion Collider (RHIC) has carried out 
systematic measurement of hadrons in $p+p$, $d+$Au, Cu$+$Cu and Au$+$Au 
collisions at $\sqrt{s_{NN}}$ =200 GeV. When compared to existing 
measurements in $p+p$ and $d+$Au, measurements in heavy-ion collisions 
suggest that particle production at high $p_T$ is affected by jet 
quenching, which is considered to be an effect of extremely dense matter 
created by the collisions \cite{Jet}. High $p_T$ suppression of $\pi^{0}$ 
and $\eta$ was measured in Cu$+$Cu and Au$+$Au \cite{ppg051, star:pi0, 
ppg080, ppg115} and the nuclear modification factors ($R_{\rm AA}$) of 
these mesons were found to be consistent with each other in $p_T$ and 
centrality. A comparison with theoretical models was first done for 
$\pi^{0}$ suppression in \cite{ppg080}, with the result that the 
suppression increases proportional to the number of participating nucleons 
as $N_{\rm part}^{2/3}$. This result is consistent with existing energy loss 
models such as the Parton Quenching Model (PQM) \cite{PQM}.


\begin{table*}[!t]
\caption{Summary of the analyzed data samples and $\omega$ meson decay channels.
Values for previously published PHENIX data (PRD83)~\cite{ppg099}
and (PRC75)~\cite{ppg064} are given for comparison.
}
\label{table:SummaryData}
\begin{ruledtabular}\begin{tabular}{lcccccc}
 Data set  & Trigger  &  \hspace{2mm} Sampled events  &  $\int Ldt$  & Threshold  
\hspace{2mm}& Decay channel  \hspace{2mm}& Reference \\ \hline
 2003 $d+$Au  &  ERT  &  5.5B  & 2.74 nb$^{-1}$  &  2.4 GeV  & 
$\omega\rightarrow \pi^{+}\pi^{-}\pi^{0}$  & {\sc PRC75}~\cite{ppg064}\\ 
\\
  &  &  &  &  2.4 GeV  & $\omega\rightarrow \pi^{0}\gamma$  & {\sc PRC75}~\cite{ppg064}  \\ 
\\
 2004 Au$+$Au &  MB  &  1.5B  &  241  $\mu$b$^{-1}$ &  N/A  & $\omega\rightarrow 
\pi^{0}\gamma$  & this work \\ 
\\
 2005 $p+p$  &  ERT  &  85B  &  3.78 pb$^{-1}$  &  0.4 GeV  & $\omega\rightarrow 
e^{+}e^{-}$  & {\sc PRD83}~\cite{ppg099} \\
  &  &  &  &  1.4 GeV  & $\omega\rightarrow \pi^{+}\pi^{-}\pi^{0}$  & {\sc PRD83}~\cite{ppg099} \\
  &  &  &  &  1.4 GeV  & $\omega\rightarrow \pi^{0}\gamma$  & {\sc PRD83}~\cite{ppg099} \\
\\
 2005 Cu$+$Cu &  MB  &  8.6B  &  3.06 pb$^{-1}$  &  N/A  & 
$\omega\rightarrow \pi^{0}\gamma$  & this work \\
  &  ERT  &  &  &  3.4 GeV  & $\omega\rightarrow \pi^{0}\gamma$  & this work \\ 
\\
 2007 Au$+$Au &  MB  &  5.1B  &  813  $\mu$b$^{-1}$ &  N/A & $\omega\rightarrow 
\pi^{0}\gamma$  & this work \\ 
\\
 2008 $d+$Au  &  ERT  &  160B  &  80  nb$^{-1}$  &  0.6/0.8 GeV  & 
$\omega\rightarrow e^{+}e^{-}$  &  this work  \\ 
  &  &  &  &  2.4 GeV  & $\omega\rightarrow \pi^{+}\pi^{-}\pi^{0}$ &  this work 
\\
  &  &  &  &  2.4 GeV  & $\omega\rightarrow \pi^{0}\gamma$ &  this work \\  
\end{tabular}\end{ruledtabular}
\end{table*}


The $\omega$ meson comprises light valence quarks similar to the 
$\pi^{0}$ and $\eta$, but has a larger mass (782 MeV) and a spin (1).  
These differences make the omega measurement an additional probe to a 
systematic study to understand mechanisms of parton energy loss and 
hadron production in the collisions. The $p_T$ dependence of the 
particle production ratio ($\omega/\pi$) and the nuclear modification 
factors ($R_{\rm AA}$) should add information about the parton energy 
loss mechanism. Furthermore, using multiple decay channels: a leptonic 
channel $\omega \rightarrow e^{+}e^{-}$ (with branching ratio 
BR=7.18$\pm$0.12$\times$10$^{-5}$) and two hadronic decay channels 
$\omega\rightarrow \pi^{+}\pi^{-}\pi^{0}$
(BR=(89.1$\pm$0.7)$\times$10$^{-2}$) and $\omega\rightarrow \pi^{0}\gamma$ 
(BR=(8.90+0.27-0.23)$\times$10$^{-2}$) \cite{PDG} extends the $p_T$ range 
by using the hadronic channels at high $p_T$ and the leptonic channel at 
low $p_T$.

Baseline measurements of the $\omega$ have been performed for $p+p$ via 
the leptonic channel \cite{ppg099}, and for the $p+p$ and $d+$Au in the 
hadronic channel \cite{ppg064, ppg055}. The $\omega/\pi^{0}$ ratio was 
found to be independent of transverse momentum and equal to 
$0.85~\pm~0.05^{\rm stat}~\pm~0.09^{\rm syst}$ in $p+p$ and 
$0.94~\pm~0.08^{\rm stat}~\pm~0.12^{\rm syst}$ in $d+$Au collisions
for $p_T>2$ GeV/$c$ \cite{ppg064}.

This article presents the first measurements of $\omega$ meson production 
in Cu$+$Cu and Au$+$Au collisions at PHENIX via the $\pi^{0}\gamma$ 
channel.  These measurements permit the study of $\omega$ suppression at 
high $p_T$. This paper also presents measurements of $\omega$ in $d+$Au 
collisions with significantly reduced uncertainties in the hadronic 
channel and a first measurement in the dielectronic channel.

\section{\label{sec:exp}Experimental Setup}

The PHENIX experiment is designed specifically to measure electromagnetic 
probes such as electrons, muons, and photons \cite{Adcox:2003zm}.  The 
detectors of the PHENIX experiment can be grouped into three categories: 
inner detectors close to the beam pipe, two central arms with 
pseudorapidity coverage of $\pm$0.35, each covering 90 degrees in 
azimuthal angle, and two muon detectors, which have 2$\pi$ azimuthal and 
pseudorapidity coverage of +(1.2--2.2) for the south muon arm and 
-(1.2--2.4) for the north muon arm. The central arms are used to measure 
the $\omega$ mesons at midrapidity.

The inner detectors are used for triggering, measurement of the 
$z$-coordinate of the collision vertex and centrality of the interactions 
with beam-beam counters (BBC) and zero degree calorimeters (ZDC). The 
central arms are capable of measuring a variety of particles by using 
Drift Chambers (DC) and Pad Chambers (PC) for tracking and momentum 
measurement of charged particles, Ring Imaging \v{C}erenkov detectors 
(RICH) for the separation of electrons up to the $\pi$ \v{C}erenkov 
threshold at 4 GeV/$c$, and an Electromagnetic Calorimeter (EMCal) for the 
measurement of spatial positions and energies of photons and electrons. 
The EMCal comprises six sectors of Lead Scintillator Calorimeter (PbSc) 
and two sectors of Lead Glass Calorimeter (PbGl). Additional details of 
the PHENIX experimental setup and performance of the detector subsystems 
can be found elsewhere \cite{NIM, ppg099}.

We used data samples collected in 2004, 2005, 2007, and 2008 as summarized 
in Table~\ref{table:SummaryData}. The data were taken using a minimum bias 
trigger (MB) and the EMC-RICH-Trigger (ERT), which is described below. The 
2003 $d+$Au data were published in \cite{ppg064} and are included here for 
comparison. The 2005 $p+p$ data were published in \cite{ppg099} and are 
used as the baseline for $R_{\rm AA}$ in $d+$Au, Cu$+$Cu and Au$+$Au. Two 
Au$+$Au data samples were taken in 2004 and 2007.  The MB trigger required 
a coincidence between the north and south BBC \cite{Allen:2003zt}. In the 
Au$+$Au data sample taken in 2004, additional coincidence between the ZDC 
and BBC was required. To enhance the statistics at high $p_T$, the ERT 
trigger was used in $p+p$, $d+$Au and Cu$+$Cu runs which required: 1) the 
event to satisfy the MB trigger conditions; 2) the presence of at least 
one high-$p_T$ electron or photon candidate in the event. For electron 
candidates the ERT trigger required a minimum energy deposit of 0.4 (0.6 
and 0.8) GeV/$c$ in a tile of $2 \times 2$ EMCal towers matched to a hit 
in the RICH in $p+p$ ($d+$Au) collisions. For the photon candidates the 
ERT trigger required a minimum energy deposit of 1.4, 2.4 and 3.4 GeV/$c$ 
in a tile of $4 \times 4$ EMCal towers in $p+p$, $d+$Au, and Cu$+$Cu 
collisions, respectively. In the $d+$Au and the Cu$+$Cu analysis, the MB 
data set was used to measure $\omega$ production up to 4 GeV/$c$ in $d+$Au 
and 6 GeV/$c$ in Cu$+$Cu; the ERT sample was used at higher $p_T$. The ERT 
trigger efficiencies measured for single photons and electrons and 
calculated for $\omega$ mesons is described in Section \ref{sec:ana:rec}.

\section{\label{sec:ana}Data Analysis}

In this section, we describe the event selection and data analysis for 
reconstructing the leptonic ($\omega\rightarrow e^{+}e^{-}$) and hadronic 
($\omega\rightarrow \pi^{+}\pi^{-}\pi^{0}$ and $\omega\rightarrow 
\pi^{0}\gamma$) decay channels of the $\omega$. Corrections applied to the 
raw data to calculate the $\omega$ meson invariant yields and systematic 
uncertainties related to the measurements are also presented.

\subsection{Event selection and basic analysis cut}

In the Run-4 PHENIX configuration, the correlation of the charge deposited 
in the BBCs with energy deposited in the ZDCs provides a determination of 
the centrality of the collisions. In the other runs, the centralities were 
only determined by using BBC. A Glauber Monte Carlo \cite{Glauber} with 
the BBC and ZDC responses was used to estimate the number of binary 
nucleon-nucleon collisions ($N_{\rm coll}$) and the number of 
participating collisions ($N_{\rm part}$) for each centrality bin 
\cite{Npart}.

Events are selected with a reconstructed $z$-vertex within 30 cm of the 
center of the interaction region. Charged tracks were required to have 
momenta in the range of $0.2<p_T<5.0$~(7.0)~GeV/$c$ for the 
$\omega\rightarrow e^{+}e^{-}$ analysis in $p+p$ ($d+$Au) \cite{ppg099} 
and $0.3<p_T<8.0$~GeV/$c$ for the $\omega\rightarrow 
\pi^{+}\pi^{-}\pi^{0}$ decay channel \cite{ppg064}. Charged particles with 
$p_T<0.2$~GeV/$c$ have a large bending angle in the axial magnetic field 
of the PHENIX central magnet \cite{Aizawa:2003zq} and most of them do not 
pass through the entire tracking system. Electrons and positrons are 
identified mainly by the \v{C}erenkov photons emitted in the RICH by 
requiring at least two photomultipliers hit in the RICH cells matched to 
the track \cite{Adcox:2003zp}. Also, matching of the energy measured for 
the charged track in the EMCal with the momentum measured in the tracking 
system, $|E/p-1|<0.5$, helps to further improve $\it{e}/\pi$ separation. 
Together the RICH and EMCal provide an $\it{e}/\pi$ rejection factor of 
about 1:10$^{4}$. Photon identification is performed by the shower shape 
criteria in the EMCal \cite{Aphecetche:2003zr}, and the energy of the selected 
$\gamma$ clusters is above 0.2 GeV.

\subsection{Leptonic analysis}

The leptonic analysis is done only in $p+p$ and $d+$Au. In case of 
$\omega\rightarrow e^{+}e^{-}$, all electrons and positrons reconstructed 
in each event are combined into pairs, resulting in signal peaks which sit 
on top of a combinatorial background in the invariant mass distribution. 
The uncorrelated part of the background is estimated with an event-mixing 
technique, which combines tracks from different events with similar event 
centrality and $z$-coordinate of the collision vertex. Details of the 
event mixing procedure are presented in \cite{ppg088}.

\begin{figure}[t]
\includegraphics[width=1.0\linewidth]{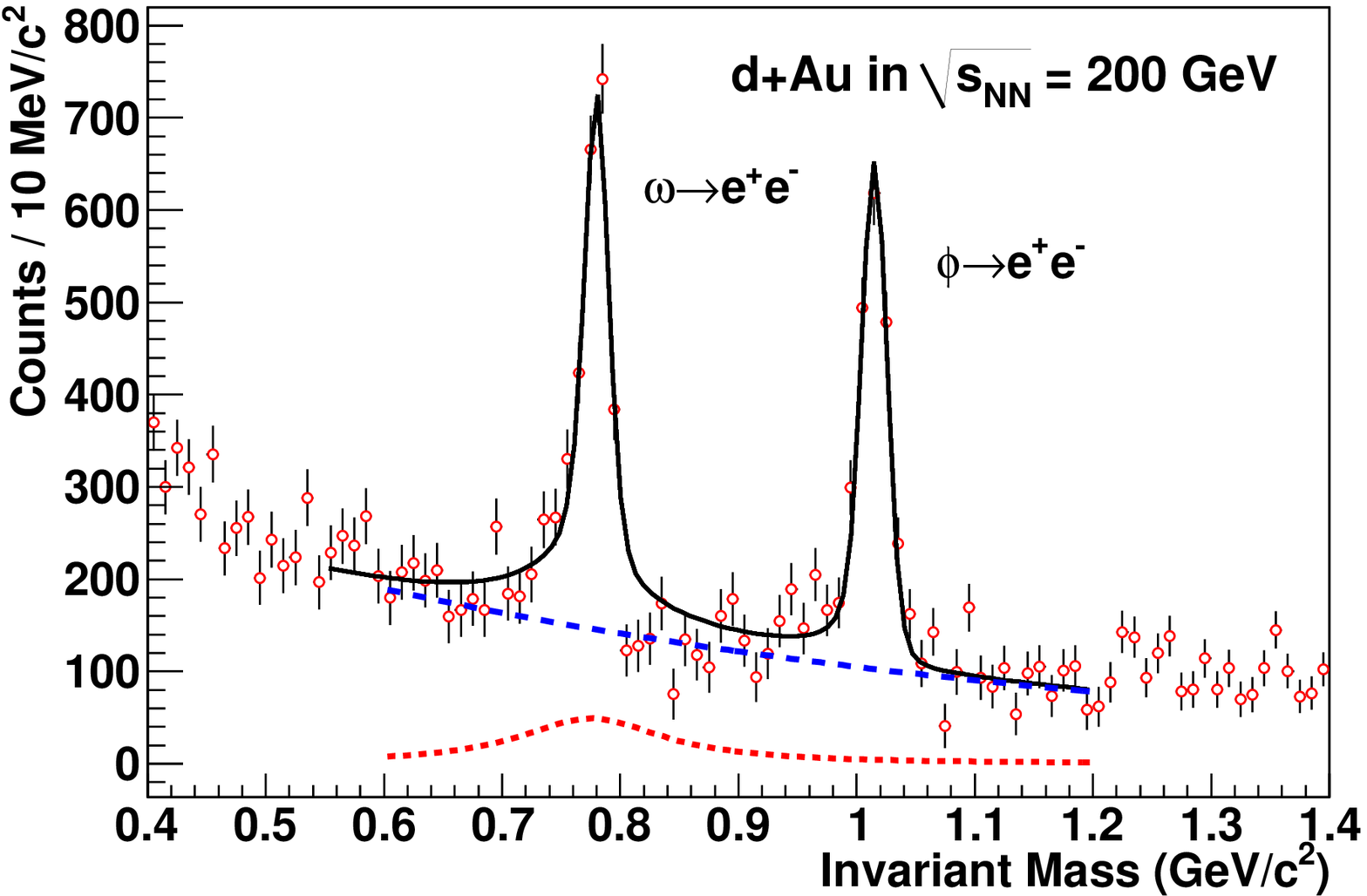}
\includegraphics[width=1.0\linewidth]{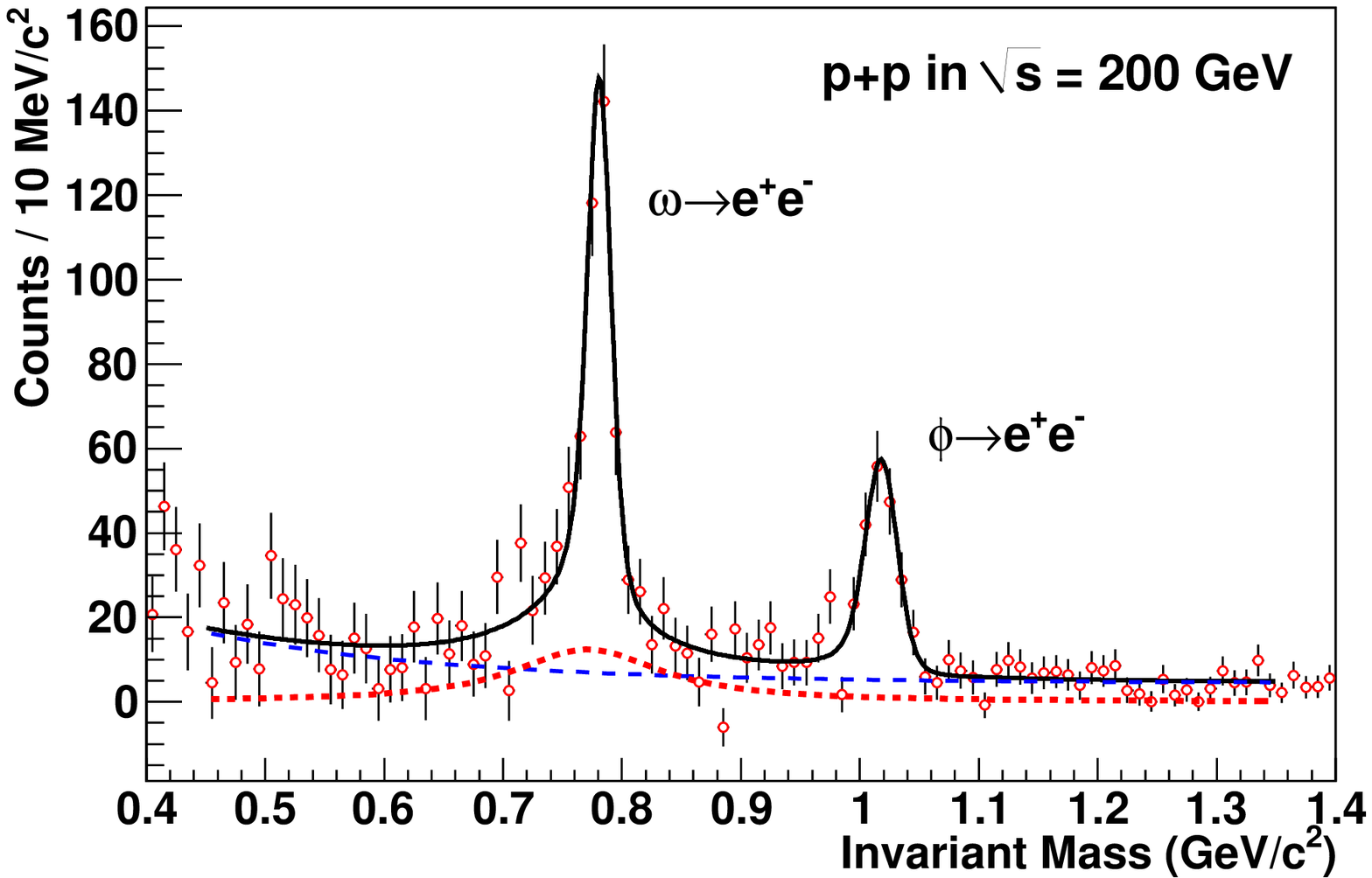}
\caption{\label{fig:massspectra_ee_pp}
Invariant mass of $e^{+}e^{-}$ pairs detected by the PHENIX central arms 
in $p+p$ collisions (left) and minimum-bias $d+$Au collisions (right) at 
$\sqrt{s_{NN}}$=200 GeV and integrated over $p_T$.  Uncorrelated 
combinatorial background is subtracted as described in the text. The 
spectrum is fit to the $\omega$ and $\phi$ resonances where the masses and 
widths are set to the PDG values; the Breit-Wigner resonance shape is 
convolved with a Gaussian to account for detector mass resolution estimated 
from simulation and then corrected for the radiative tail. The $\rho$ 
contribution is shown as the dotted line with an assumption that the yield 
is the same as that of the $\omega$.  The residual continuum component is 
estimated by a polynomial fit as shown by the dashed line.
}
\end{figure}

Figure \ref{fig:massspectra_ee_pp} shows invariant mass spectra of 
$e^{+}e^{-}$ pairs in $p+p$ and $d+$Au collisions at $\sqrt{s_{NN}}$ = 200 
GeV after subtraction of combinatorial background as described above. The 
solid lines show the global fits which include: (1) contributions from 
$\omega$, $\rho$ and $\phi$ mesons approximated with Breit-Wigner 
functions convolved with Gaussian distributions to account for the 
detector mass resolution; masses and widths of the $\omega$, $\rho$ and 
$\phi$ are fixed to the PDG values; the $\rho$ component is calculated 
assuming that $\omega$ and $\rho$ have the same yield and vacuum branching 
ratios; (2) other correlated residual background, which is dominated by a 
contribution from jets, is approximated by a second order polynomial 
function.  The detector resolution, which is determined from simulations, 
is found to be dependent on mass and momentum and varies from 6 MeV/$c^2$ 
to 18 MeV/$c^2$.

The $\omega$ yield is determined by counting bin contents in a 3 $\sigma$ 
width (derived from the fitting) and subtracting the polynomial 
background.  An associated systematic uncertainty from the raw yield 
extraction is calculated by varying the background normalization, fitting 
functions, range and counting methods. The estimated value is 4--15\% in 
$p+p$ \cite{ppg099} and 8--15\% in $d+$Au collisions.

\subsection{Hadronic analysis}

In the $\omega\rightarrow \pi^{+}\pi^{-}\pi^{0}$ and 
$\omega\rightarrow \pi^{0}\gamma$ channels, the first analysis step is 
to reconstruct $\pi^{0}$ mesons by combining pairs of photons 
reconstructed in an event. Then the mass and width of the $\pi^{0}$ 
peak in the invariant mass distribution of photon pairs are 
parametrized as a function of transverse momentum. The 1 $\sigma$ 
width of the $\pi^{0}$ peak varies from 13 MeV/$c^2$ to 9 MeV/$c^2$ as 
$p_T$ increases from 1 GeV/$c$ to 4 GeV/$c$ and is determined by the 
EMCal energy resolution. A pair of photons is selected as a $\pi^{0}$ 
candidate if its invariant mass is within 2 $\sigma$ of the 
reconstructed $\pi^{0}$ mass. In Cu$+$Cu and Au$+$Au, an additional 
asymmetry cut for $\pi^{0}$ candidates is used to reduce combinatorial 
background, 
$\alpha = |E_{\gamma_{1}}-E_{\gamma_{2}}|/|E_{\gamma_{1}}+E_{\gamma_{2}}|<0.8$. 
Selected $\pi^{0}$ candidates, which include true $\pi^{0}$s and 
combinatorial background are combined either with the third photon 
with energy $E_{\gamma}>$ 1.0 GeV/$c$ for the 
$\omega\rightarrow \pi^{0}\gamma$ or with a pair of opposite-sign 
charged tracks for the $\omega\rightarrow \pi^{+}\pi^{-}\pi^{0}$ 
decay.

In the $p+p$ and $d+$Au analysis, the $\omega$ meson raw yields are 
extracted by fitting the $p_T$ slices of the invariant mass distribution 
with a combination of a Gaussian for the signal and a second order 
polynomial for the background. The width and mass of the reconstructed 
$\omega$ mesons were found to be in good agreement with values expected 
from simulation. Details of these analyses are described in \cite{ppg055}.

\begin{figure}[t]
\includegraphics[width=1.0\linewidth]{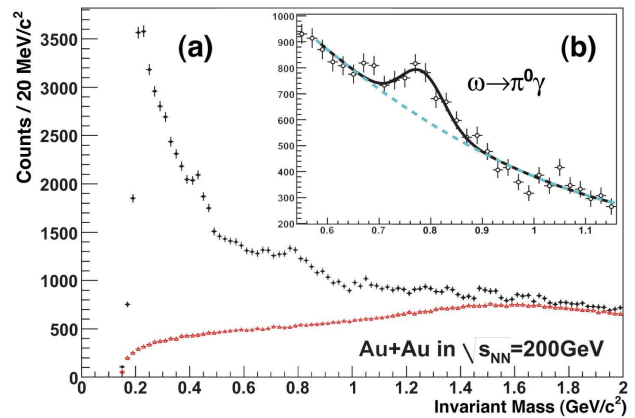}
\caption{\label{fig:massspectra_AuAu} 
(a) Invariant mass and scaled mixed background distributions for 
$\pi^{0}\gamma$ decay at $7<p_T<12$ GeV/$c$ in Au$+$Au collisions
(b) Invariant mass distribution after subtraction of scaled background.
}
\end{figure}

In the Cu$+$Cu and Au$+$Au analysis, only the $\omega\rightarrow 
\pi^{0}\gamma$ channel was analyzed due to high combinatorial background 
in the $\omega\rightarrow \pi^{+}\pi^{-}\pi^{0}$ channel. The 
uncorrelated combinatorial background was estimated using an event 
mixing technique where the third photon in the $\pi^{0}\gamma$ decay was 
taken from the different events with a similar centrality and $z$-vertex. 
For every $p_T$ bin the calculated background was normalized to match 
the integral of the foreground at an invariant mass $M_{inv}>1.75$ 
GeV/$c^2$ and then subtracted. An example of the invariant mass 
distribution and normalized background distributions is shown in 
Fig.~\ref{fig:massspectra_AuAu}(a) with the invariant mass distribution 
after subtraction shown in Fig.~\ref{fig:massspectra_AuAu}(b).  The 
resulting invariant mass distribution contains residual background from 
correlated particles: the background contributions are from 
$K_s\rightarrow \pi^{0}\pi^{0}$ decays, and $\pi^{0}$ and $\eta$, where 
one of the photons from $\pi^{0}$($\eta$)$\rightarrow \gamma\gamma$ 
decay creates a fake $\pi^{0}$ candidate for the $\omega\rightarrow 
\pi^{0}\gamma$ reconstruction. The $\omega\rightarrow \pi^{0}\gamma$ 
peak is further enhanced by a mixed background subtraction. Finally, raw 
yields of $\omega$ are extracted by fitting the spectra with a 
combination of a Gaussian and a polynomial. The width of the Gaussian is 
limited $\pm$ 1 MeV/$c^2$ in the fit to the data as determined from 
simulation. The $\omega$ yield is calculated as an integral of the 
Gaussian.

Systematic uncertainties associated with the raw yield extraction are 
evaluated using different fitting functions and ranges, different 
counting methods and kinematic cuts, varying the EMC resolution in 
simulation, and applying different limits for the width of $\omega$ 
peaks in fits to data. The estimated value is 13--35\% in Cu$+$Cu and 
20--35\% in Au$+$Au collisions.


\subsection{\label{sec:ana:rec}Reconstruction efficiencies}

The reconstruction efficiency of the $\omega$ is determined using a 
{\sc geant} simulation of the PHENIX detector tuned to reproduce the 
performance of the detector subsystems. The $\omega$ mesons are 
generated and decayed into corresponding decay channels, and 
reconstructed with the same analysis chain as the real data. The 
generated $\omega$ spectra were weighted to match the measured 
particle spectra. It was verified that the simulated positions and 
widths of the reconstructed particle peaks are consistent with the 
values measured in real data.

\begin{figure}[t]
\includegraphics[width=0.98\linewidth]{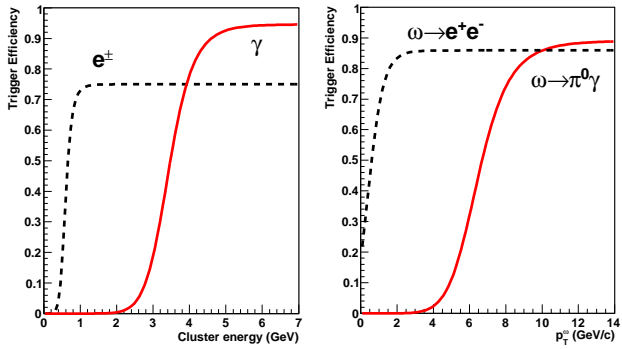}
\caption{\label{fig:ERTEff} 
Typical ERT trigger efficiency. Left: trigger efficiencies for single 
electrons (0.6 GeV threshold) and photons (3.4 GeV threshold). Right: 
trigger efficiencies for $\omega\rightarrow e^{+}e^{-}$ and 
$\omega\rightarrow \pi^{0}\gamma$ using corresponding triggered 
electrons/photons.
}
\end{figure}

\begin{figure}[h]
\includegraphics[width=0.97\linewidth]{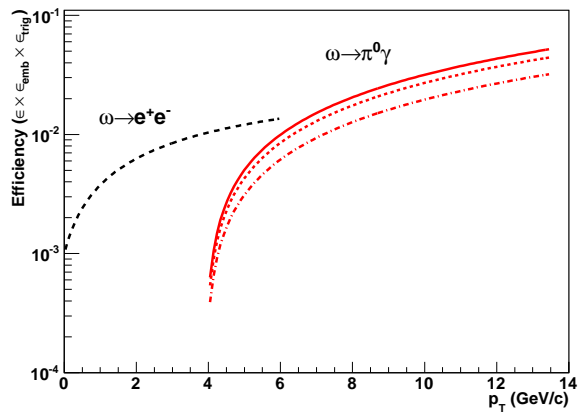}
\caption{\label{fig:totaleff} 
Typical reconstruction efficiencies for $\omega\rightarrow e^{+}e^{-}$ and 
$\omega\rightarrow\pi^{0}\gamma$. The curve for 
$\omega\rightarrow\pi^{0}\gamma$ includes the embedding efficiency in 
Au$+$Au collisions: solid, dotted, and dot-dashed lines, are respectively 
for 60--92\% , 20--60\%, and 0--20\% centrality.
}
\end{figure}

The reconstruction efficiency is divided into three components: 
$\epsilon$, $\epsilon_{\rm trig}$, and $\epsilon_{\rm emb}$.  
The efficiency $\epsilon$ is the reconstruction efficiency for minimum 
bias events in a low occupancy environment, like in $p+p$ and $d+$Au 
collisions. This efficiency accounts for the limited geometrical 
acceptance, resolution and efficiencies of the detector subsystems as 
well as for analysis cuts. When a selective ERT trigger is used, an 
additional trigger efficiency factor, $\epsilon_{\rm trig}$, is 
applied.  This factor measures the efficiency of the ERT trigger 
logic. For higher multiplicity collisions, one needs to account for 
the loss of efficiency from increased detector occupancy: this is 
measured through the embedding efficiency $\epsilon_{\rm emb}$.  A 
measured raw yield then needs to be corrected for the total efficiency 
$\epsilon \times \epsilon_{\rm emb} \times \epsilon_{\rm trig}$, 
depending on the collision, centrality, and trigger involved.

The ERT data sample was used to measure dielectron and hadronic decay 
channels of the $\omega$ at high $p_T$ in $p+p$, $d+$Au and Cu$+$Cu. 
The threshold settings for ERT are described in Section \ref{sec:exp}. 
The single particle ERT efficiency was measured by dividing the energy 
spectra of gamma clusters or electrons that fired the ERT trigger by 
the energy spectra of all clusters or electrons in the minimum bias 
data sample. A typical example of the ERT trigger efficiencies for 
single electrons and single photons as a function of cluster energy is 
shown on the left in Fig.~\ref{fig:ERTEff}. The level of saturation of 
trigger efficiency curves is below 100\% because of inactive areas of 
the ERT and the RICH detectors.

The ERT efficiencies for the $\omega$ meson in both the leptonic and 
hadronic decay modes were evaluated with the help of a Monte-Carlo 
simulation. For all fully reconstructed $\omega$ mesons, the 
calculated single photon or electron ERT efficiency curves were used 
to calculate the probability that one of the particles in the final 
state fires the ERT trigger. Corresponding trigger efficiencies for 
$\omega\rightarrow e^{+}e^{-}$ and $\omega\rightarrow \pi^{0}\gamma$ 
are shown on the right in Fig.~\ref{fig:ERTEff}. More detailed 
descriptions are presented in \cite{ppg055, ppg099}.

Figure \ref{fig:totaleff} shows typical reconstruction efficiencies 
$\epsilon$ for $\omega\rightarrow e^{+}e^{-}$ and $\omega\rightarrow 
\pi^{0}\gamma$. In the case of Cu$+$Cu and Au$+$Au collisions, an 
additional efficiency correction $\epsilon_{\rm emb}$ due to cluster 
overlap in high multiplicity environment must be applied. In most 
central Au$+$Au events, the EMCal typically detects more than 300 
clusters corresponding to a detector occupancy of $\sim$10\%. To 
estimate the corresponding loss in efficiency, the simulated $\omega$ 
decays are embedded into real A$+$A events and analyzed. The merging 
effect results in $\sim$40\% loss of reconstruction efficiency in 
0--20\% central Au$+$Au collisions, $\sim$15\% loss in 0--20\% central 
Cu$+$Cu collisions and is almost negligible in peripheral collisions. 
The reconstruction efficiencies derived for Au$+$Au collisions at 
different centralities are shown in Fig.~\ref{fig:totaleff}. Finally, 
in each bin we apply also a correction factor \cite{ppg099} to replace 
the average value of the yield in the analyzed $p_T$ bin by the value 
of the yield in the middle of the bin.

\subsection{Calculation of invariant yields}

In $p+p$ and minimum bias $d+$Au collisions, the invariant yield is 
related to the invariant cross section as: 
 \begin{equation} 
E\frac{d^{3}\sigma}{dp^{3}} = 
\sigma^{inel}_{pp}(\sigma^{inel}_{dAu}) \times 
\frac{1}{2\pi p_T}\frac{d^{2}N}{dp_Tdy}, 
 \end{equation}
where $\sigma^{inel}_{pp}$ and $\sigma^{inel}_{dAu}$ are the total 
inelastic cross section, 42.2 mb and 2260 mb respectively.

For a given centrality bin the invariant yields as a function of $p_T$ 
(invariant transverse momentum) are determined from:
 \begin{widetext} 
 \begin{equation}
\frac{1}{2\pi p_T}\frac{d^{2}N_{\rm cent}}{dp_Tdy}\equiv 
\frac{1}{2\pi p_TN^{\rm evt}_{\rm 
cent}}\frac{1}{BR}\frac{1}{\epsilon(p_T)\epsilon_{\rm 
emb}(p_T,cent)\epsilon_{\rm trig}(p_T)}
\frac{N(\Delta p_T,cent)}{\Delta p_T\Delta y},
 \end{equation}	
 \end{widetext}
where $N^{\rm evt}_{\rm cent}$ is the number of events for a given 
centrality bin, $N(\Delta p_T,cent)$ is the raw yield of $\omega$ for 
each $p_T$ and centrality bin, $\epsilon(p_T)$, $\epsilon_{\rm 
emb}(p_T,cent)$ and $\epsilon_{\rm trig}(p_T)$ are, as previously 
defined, reconstruction efficiency, embedding efficiency and trigger 
efficiency, respectively. The trigger efficiency is applied only for 
the analyses using the ERT data set. BR is the decay branching ratio 
from \cite{PDG},
(89.2$\pm$ 0.7$\times 10^{-2}$) for $\omega\rightarrow \pi^{+}\pi^{-}\pi^{0}$ ,
(8.90$\pm$ 0.27$\times 10^{-2}$) for $\omega\rightarrow\pi^{0}\gamma$ and
(7.16$\pm$ 0.12$\times 10^{-5}$) for $\omega\rightarrow e^{+}e^{-}$.

\subsection{Systematic uncertainties}

In addition to uncertainties related to the raw yield extraction 
described in the corresponding analysis sections, other sources of the 
uncertainties should also be taken into account. Uncertainties of the 
ERT trigger efficiency and acceptance corrections were estimated by 
varying the analysis cuts, energy and momentum scales of the EMCal and 
DC by $\sim$1\% \cite{ppg099}. Uncertainties of detector response 
(mainly from the RICH for electron analysis and from the EMCal for 
hadronic analysis) are estimated by changing particle identification 
criteria in the analysis. A summary of assigned systematic 
uncertainties is listed in
Table~\ref{table:SummaryError1} for $\omega\rightarrow e^{+}e^{-}$ in 
$p+p$ and $d+$Au and in Table \ref{table:SummaryError2} for 
$\omega\rightarrow \pi^{0}\gamma$ in Cu$+$Cu and Au$+$Au. Those are 
classified into three types: Type A is $p_T$-uncorrelated, Type B is 
$p_T$-correlated and Type C is the overall normalization uncertainty. 
Total uncertainties for $\omega\rightarrow e^{+}e^{-}$ are 16--24\% in 
$p+p$~\cite{ppg099} and 19--26\% in $d+$Au.  The total uncertainties for 
$\omega\rightarrow \pi^{0}\gamma$ are 15--37\% in Cu$+$Cu and 21--37\% in 
Au$+$Au. Uncertainties for $\omega\rightarrow\pi^{0}\pi^{+}\pi^{0}$ 
analysis are 7--20\% in $p+p$ and 10--15\% in $d+$Au, as described in 
\cite{ppg064}.

\section{\label{sec:res}Results}

\subsection{\label{subsec:inv}Invariant transverse momentum spectra}

\begin{figure*}[th]
\includegraphics[width=0.85\linewidth]{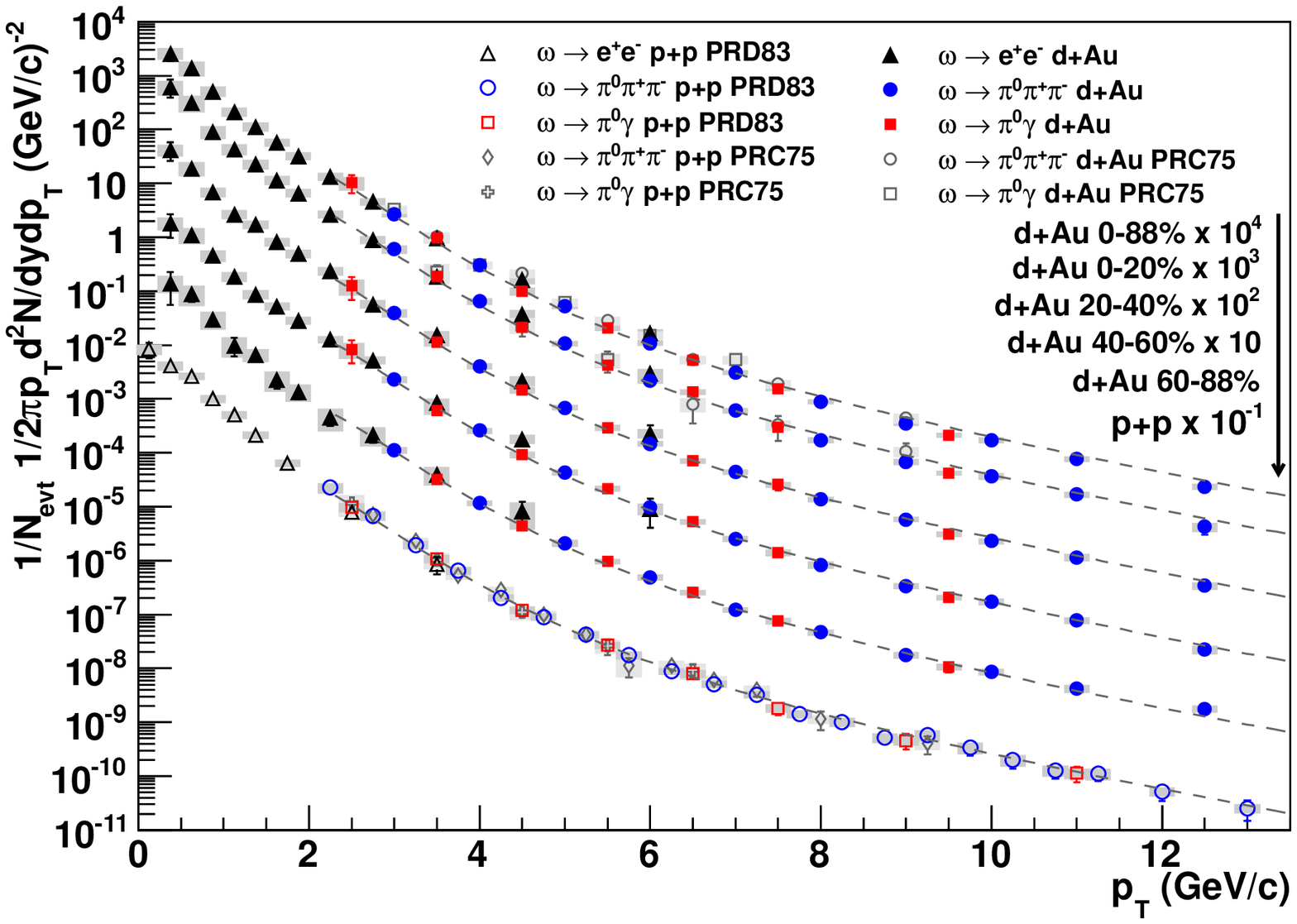}
\caption{\label{fig:ptspectra_pp_dAupT} 
Invariant transverse momentum spectra of $\omega$ production in $p+p$ 
and $d+$Au collisions at $\sqrt{s}$=200 GeV. The dashed lines 
represent fits to $p+p$ results and those are scaled by the 
corresponding number of binary collisions for $d+$Au . The previously 
published PHENIX data (PRD83)~\cite{ppg099} and 
$\pi^{0}$~(PRC75)~\cite{ppg064} are shown for comparison.
}
\includegraphics[width=0.8\linewidth]{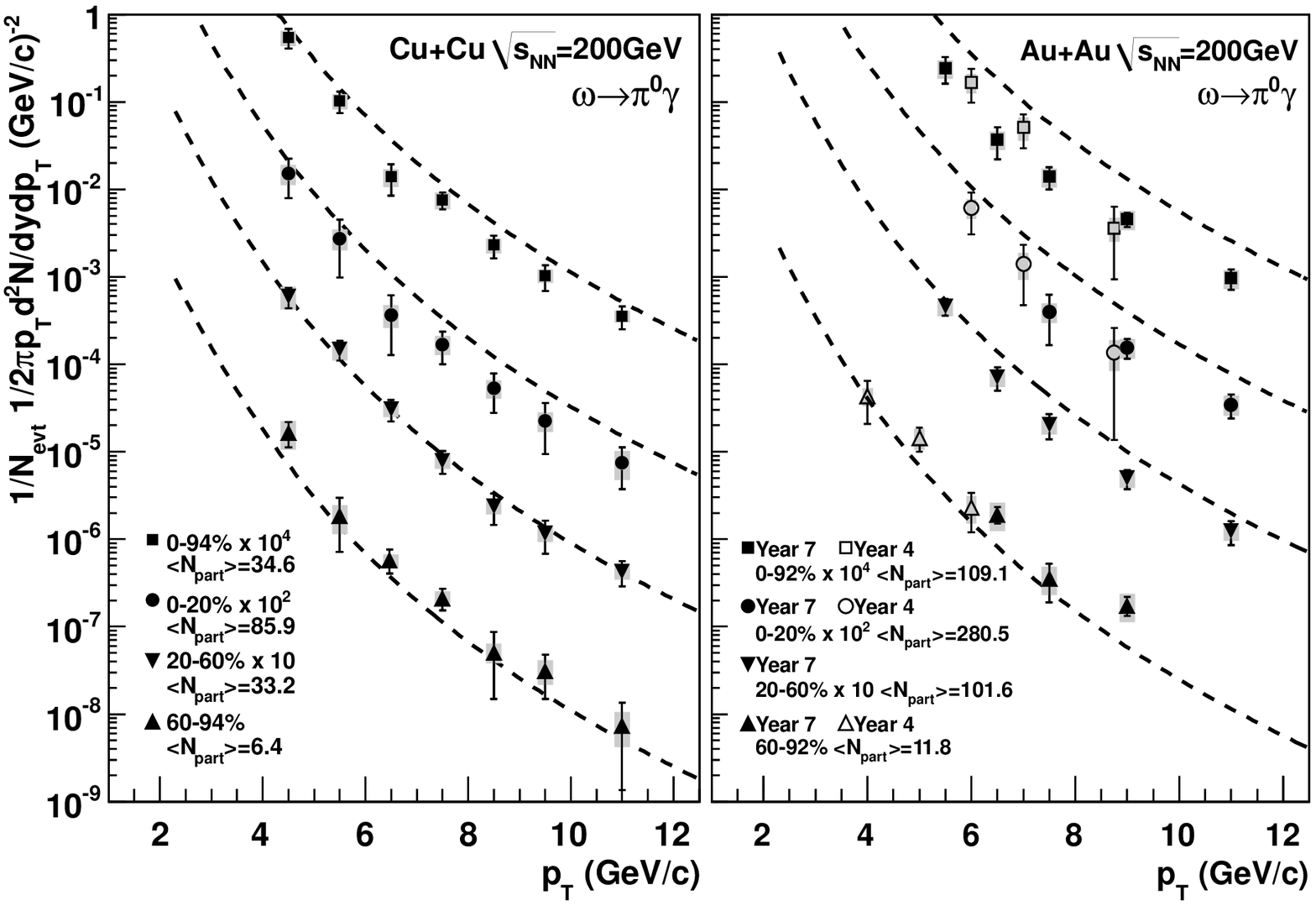}
\caption{\label{fig:ptspectra_AuAupT} 
Invariant transverse momentum spectra of $\omega$ production in 
Cu$+$Cu (left) and Au$+$Au (right) collisions from the 
$\omega\rightarrow \pi^{0}\gamma$ decay channel for three centrality 
bins and minimum bias. The dashed lines are the $p+p$ results scaled 
by the corresponding number of binary collisions. The Cu$+$Cu (left) 
data was recorded in 2005. For Au$+$Au (right) the Year 7 and Year 4 
refer to data taken in 2007 and 2004, respectively.
}
\end{figure*}

Figure \ref{fig:ptspectra_pp_dAupT} presents the invariant transverse 
momentum spectra measured for the $\omega$ in $p+p$ and $d+$Au at 
$\sqrt{s}$=200 GeV. Previously published results are shown with open 
markers \cite{ppg064}. Results for different decay channels and data 
samples agree within uncertainties in the overlap region. The dashed 
curves in Fig.~\ref{fig:ptspectra_pp_dAupT} are fixed on $p+p$ results at 
$p_T>$ 2 GeV/$c$ and then scaled by the number of binary nucleon-nucleon 
collisions ($N_{\rm coll}$) estimated using Glauber Monte-Carlo simulation 
\cite{Glauber} for $d+$Au results.

\begin{table}[b]
\caption{Summary of assigned systematic uncertainties of $\omega\rightarrow 
e^{+}e^{-}$ 
in $p+p$ and $d+$Au analysis.}
\label{table:SummaryError1}
\begin{ruledtabular}\begin{tabular}{lcc} 
Source  \hspace{5mm} &  $p+p$  \hspace{3mm} & $d+$Au \\ \hline
peak extraction  \hspace{5mm} &  4--15\%(A) \hspace{3mm} & 8.4--24.1\%(A) \\
ERT efficiency  \hspace{5mm} &  1--3\%(B)  \hspace{3mm} & 1--7\%(B) \\
BBC cross section  \hspace{5mm} &  9.7\%(C)  \hspace{3mm} & 7.9\%(C) \\
momentum scale  \hspace{5mm} &  2--11\%(B)  \hspace{3mm} & 1.2--5.3\%(B) \\
acceptance correction  \hspace{5mm} &  5\%(B)  \hspace{3mm} & 7\%(B) \\
electron ID  \hspace{5mm} &  \multicolumn{2}{c}{10\%(B)} \\
branching ratio  \hspace{5mm} &  \multicolumn{2}{c}{1.7\%(C)} \\
\end{tabular}\end{ruledtabular}

\caption{Summary of assigned systematic uncertainties of $\omega\rightarrow 
\pi^{0}\gamma$ 
in  Cu$+$Cu and  Au$+$Au analysis.}
\label{table:SummaryError2}
\begin{ruledtabular}\begin{tabular}{lcc} 
Source  \hspace{5mm} &  Cu$+$Cu  \hspace{3mm} &  Au$+$Au \\ \hline
peak extraction  \hspace{5mm} &  13--35\%(A) \hspace{3mm} & 20.1--34.5\%(A) \\
ERT efficiency  \hspace{5mm} &  3--4\%(B)  \hspace{3mm} & N/A \\
energy scale  \hspace{5mm} &  \multicolumn{2}{c}{4--7\%(B)} \\
energy resolution  \hspace{5mm} &  \multicolumn{2}{c}{2--3\%(B)} \\
acceptance correction  \hspace{5mm} &  \multicolumn{2}{c}{3--6\%(B)} \\
conversion  \hspace{5mm} &  \multicolumn{2}{c}{4.5\%(C)} \\
branching ratio  \hspace{5mm} &  \multicolumn{2}{c}{3.4\%(C)} \\
\end{tabular}\end{ruledtabular}
\end{table}

Invariant transverse momentum spectra measured for the $\omega$ meson in 
Cu$+$Cu and Au$+$Au collisions at $\sqrt{s_{NN}}$=200 GeV are shown in 
Fig.~\ref{fig:ptspectra_AuAupT}. Measurements were performed only in the 
$\omega\rightarrow \pi^{0}\gamma$ channel. Results are presented for three 
centrality bins: 0--20\%, 20--60\%, 60--92\% (60--94\% in Cu$+$Cu) and 
minimum bias collisions. The dashed lines represent $N_{\rm coll}$ scaled 
fits to $p+p$ results, where $N_{\rm coll}$ values were taken from 
\cite{Npart}. The results show that in peripheral heavy ion collisions 
$\omega$ production generally follows binary scaling, while in midcentral 
and central collisions, production of $\omega$ mesons is suppressed at high 
$p_T$. Such behavior is similar to one previously observed for other light 
mesons \cite{ppg080, ppg050} and can be attributed to medium induced 
effects.

\subsection{\label{subsec:ratio}$\omega/\pi$ ratio}

Measurement of $\omega$ production can be used to study the relative 
production of vector and pseudoscalar mesons consisting of the same 
valence quarks, i.e. $\omega/\pi$ ratio as a function of transverse 
momentum. In calculating the $\omega/\pi$ ratio, the same methodology from 
\cite{ppg030, ppg080, ppg084} for the $\pi^{+}$/$\pi^{-}$ and $\pi^{0}$ 
was used. The charged pion results, $(\pi^{+}+\pi^{-})/2$, were used to 
extend neutral pion measurements at the lower limit of the $p_T$ range 
from 1 to 0.2 GeV/$c$. To produce the average pion spectrum in $p+p$ 
\cite{ppg030} and $d+$Au collisions \cite{ppg044}, we simultaneously fit 
$(\pi^{+}+\pi^{-})/2$ and $\pi^{0}$ spectra with the modified Hagedorn 
function \cite{ppg088}.  Inclusion of the charged pion spectrum in the fit 
has a small effect in the 1--2 GeV/$c$ overlap region, smaller than 5\% 
compared to fitting neutral pions alone. The resulting fitted pion 
distributions are used to calculate $\omega/\pi$ ratios for $p+p$ and 
$d+$Au. Uncertainties for the fit values are evaluated by taking into 
account statistical and systematic uncertainties of the experimental 
points as described in \cite{ppg099,ppg079}.

\begin{figure}[b]
\includegraphics[width=1.0\linewidth]{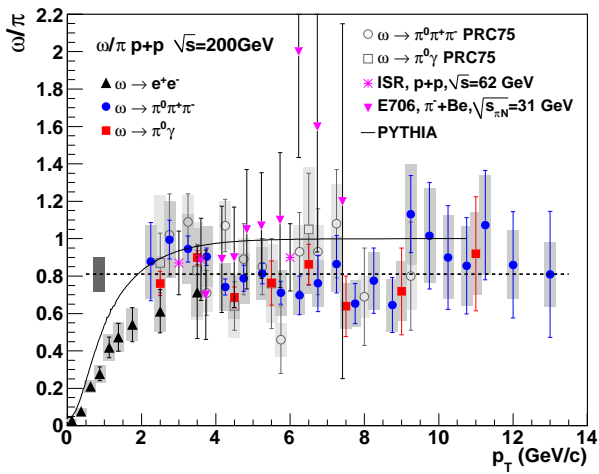}
\caption{\label{fig:omegapi0ratio} 
Measured $\omega/\pi$ ratio as a function of $p_T$ in $p+p$ collisions 
at $\sqrt{s}$=200 GeV.  (dashed line) Fit of a constant value 
to data points at $p_T>$ 2 GeV/$c$.  The fit result is 
$0.81~\pm~0.02^{\rm stat}~\pm~0.09^{\rm syst}$. 
(gray box) The overall error of the fitting. 
(solid line) The {\sc pythia} prediction \cite{PYTHIA} for $p+p$ 
at $\sqrt{s}$=200 GeV.  Previously published PHENIX results 
(PRC75)~\cite{ppg064} and other lower energy experiments at 
$\sqrt{s_{\pi N}} = 31$ GeV~(E706)~\cite{E706} and $\sqrt{s} = 62$ 
GeV~(ISR)~\cite{ISR} are shown for comparison.
}
\end{figure}

Figure~\ref{fig:omegapi0ratio} presents the $\omega/\pi$ ratio measured in 
$p+p$ collisions at $\sqrt{s}$=200 GeV as a function of transverse 
momentum. Open markers show our previous measurements of the $\omega/\pi$ 
ratio \cite{ppg064}. One can see good agreement between previous results 
and this measurement. For completeness we also present similar 
measurements performed in lower energy experiments: $\pi+$Be at 
$\sqrt{s_{NN}} = 31$ GeV (E706 \cite{E706}), $p+p$ at $\sqrt{s}$=62 GeV 
(ISR \cite{ISR}). Please note that the branching ratio for the 
$\omega\rightarrow \pi^{0}\gamma$ decay was set equal to (8.8$\pm$0.5)\%, 
which is 6\% different from the latest PDG value of (8.28$\pm$0.28)\%. 
Within measurement uncertainties the $\omega/\pi$ ratio in hadronic 
interactions is energy independent at high $p_T$.

\begin{figure*}[th]
\begin{minipage}{0.48\linewidth}
\includegraphics[width=0.99\linewidth]{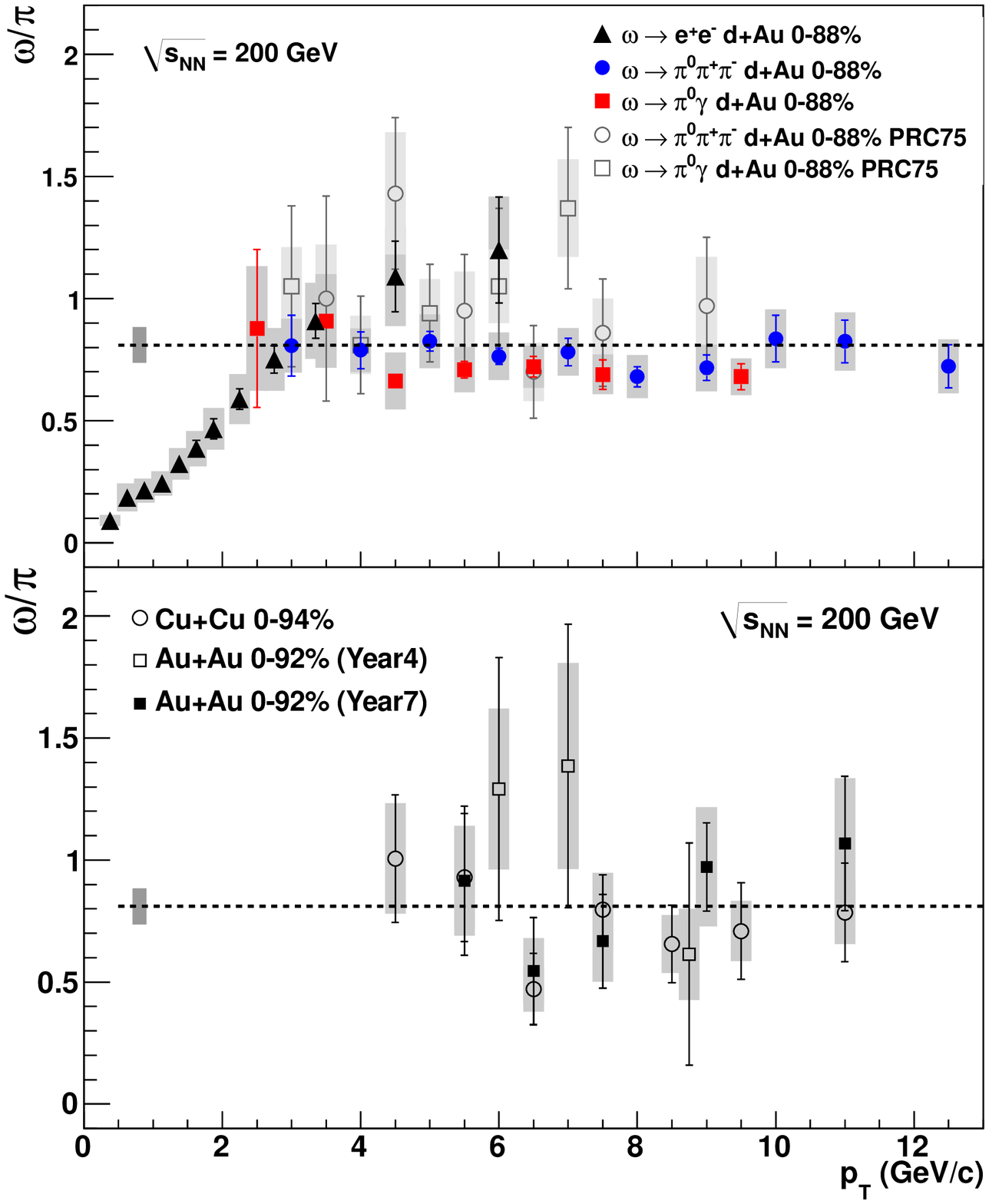}
\caption{\label{fig:omegapi0ratioA} 
Top: $\omega$/$\pi$ ratio versus transverse momentum in $d+$Au 
(0--88\%) for $\omega \rightarrow e^{+}e^{-}$, $\pi^{0}\pi^{+}\pi^{-}$ 
and $\pi^{0}\gamma$. Bottom: $\omega$/$\pi$ ratio versus transverse 
momentum in Cu$+$Cu (0--94\%) for $\omega \rightarrow \pi^{0}\gamma$ 
and in Au$+$Au (0--92\%) for $\omega \rightarrow \pi^{0}\gamma$. The 
dashed lines and boxes are a fit of a constant value to the data 
points at $p_T>$ 2 GeV/$c$ in $p+p$ (Fit result: 
$0.81~\pm~0.02^{\rm stat}~\pm~0.09^{\rm syst}$). The previously 
published data (PRC75)~\cite{ppg064} are shown for comparison.
}
\end{minipage}%
\hspace{0.5cm}
\begin{minipage}{0.48\linewidth}
\includegraphics[width=0.9\linewidth]{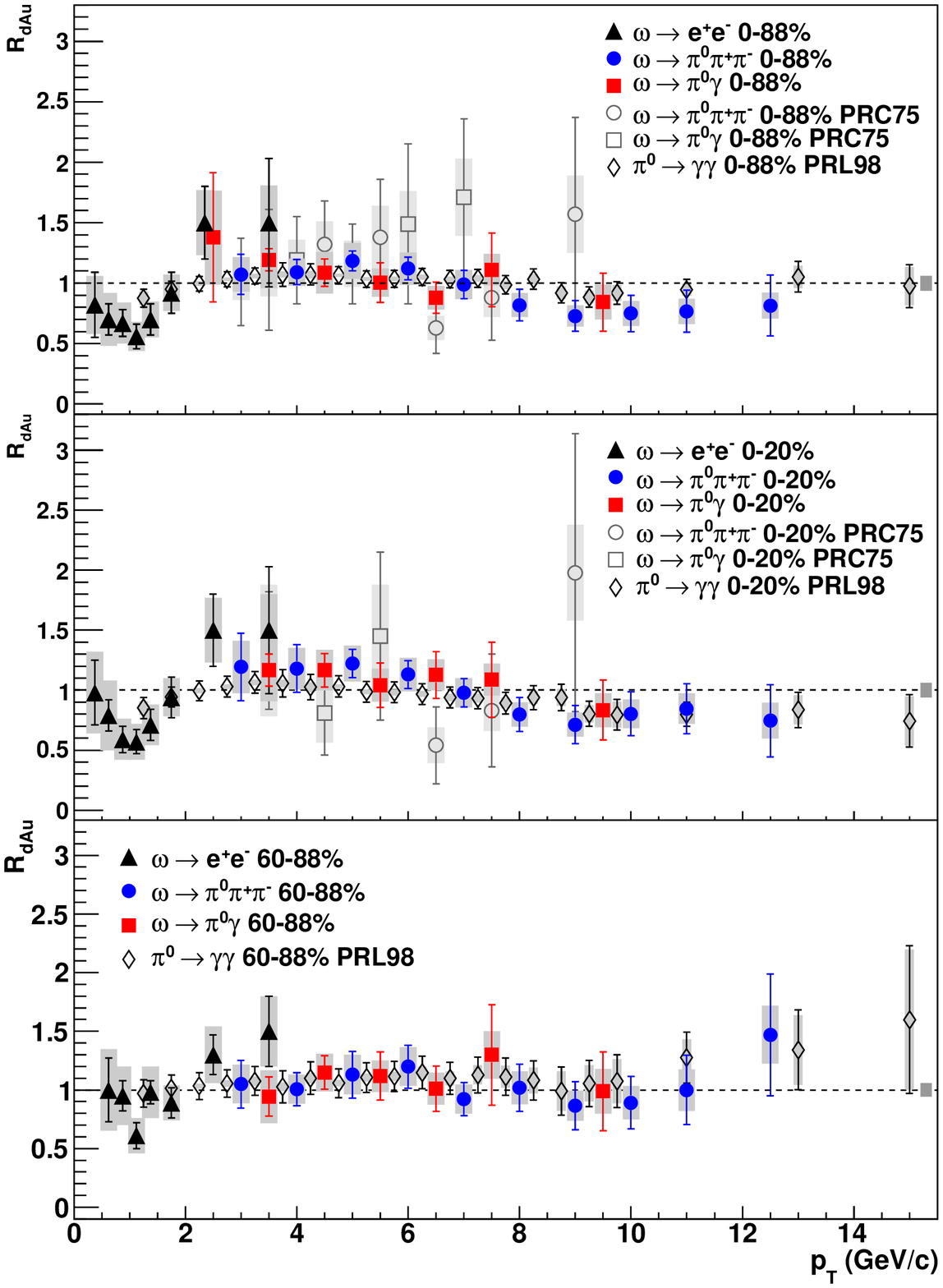}
\caption{\label{fig:omegaRdAu}
Nuclear modification factor, $R_{d{\rm Au}}$, measured for the 
$\omega$ in 0--88, 0--20, and 60--88\% centrality bins in $d+$Au 
collisions at $\sqrt{s}$=200 GeV. The box at the right edge of the 
constant fit line shows the uncertainty of the fit.  The previously 
published data for $\omega$~(PRC75)~\cite{ppg064} and 
$\pi^{0}$~(PRL98)~\cite{ppg044} are shown for comparison.
}
\end{minipage}%
\vspace{0.3cm}
\begin{minipage}{0.9\linewidth}
\includegraphics[width=0.7\linewidth]{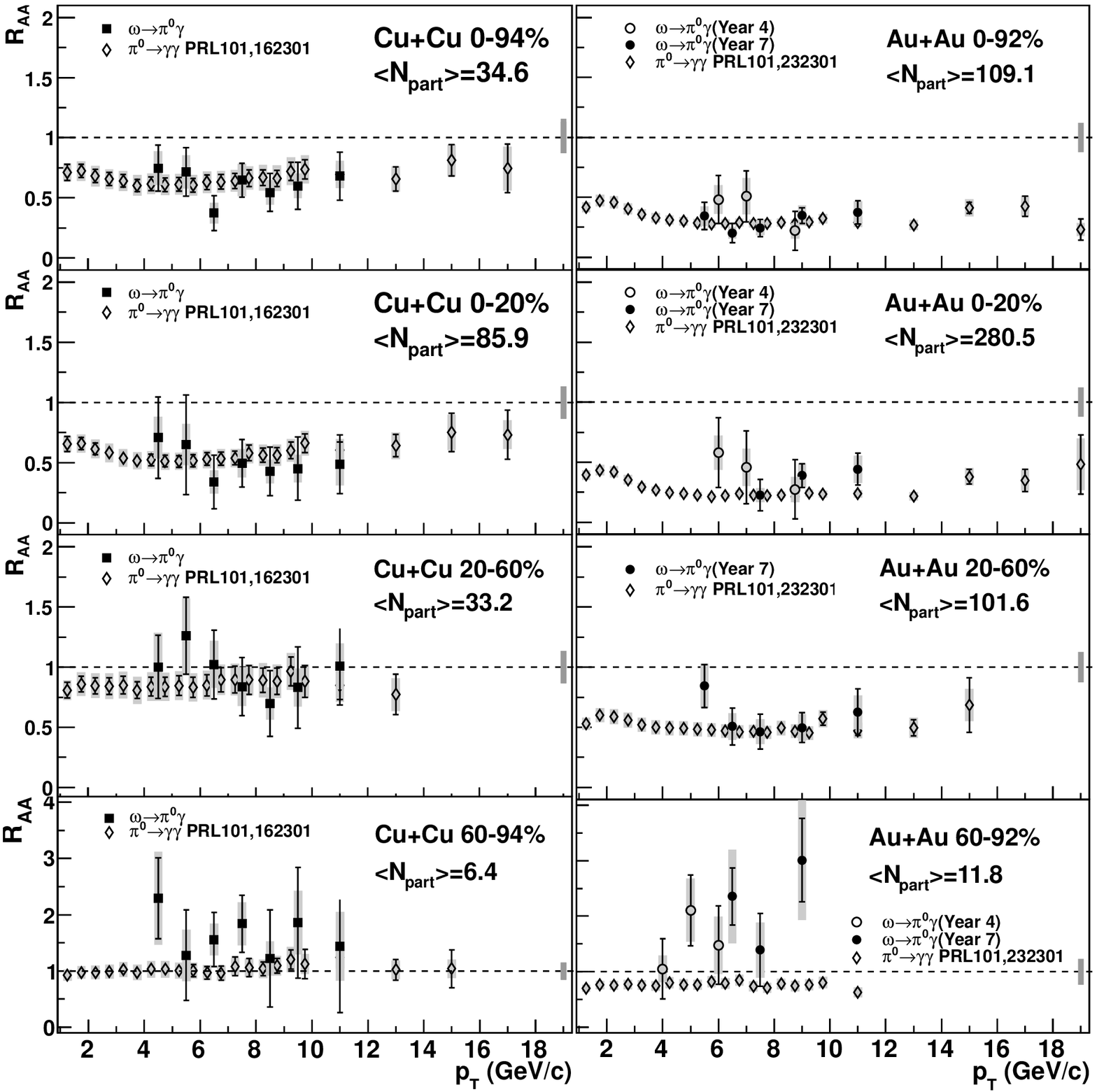}
\caption{\label{fig:omegaRAA}
$R_{\rm AA}$ of the $\omega$ in Cu$+$Cu (left) and Au$+$Au (right) 
collisions from the $\omega\rightarrow\pi^{0}\gamma$ decay channel for 
three centrality bins and minimum bias. The uncertainty in the 
determination of $p+p$ scaling is shown as a box on the left in each 
plot. Rhombuses in each plot are $R_{\rm AA}$ of $\pi^{0}$ in Cu$+$Cu 
\cite{ppg084} and Au$+$Au \cite{ppg080} shown as a comparison.
}
\end{minipage}%
\end{figure*}

A linear fit to the ratio at $p_T>$ 2 GeV/$c$ gives a value of the 
linear coefficient consistent with zero within less then one standard 
deviation ($-0.013~\pm~0.009^{\rm stat}~\pm~0.014^{\rm syst}$) 
indicating no significant $p_T$ dependence of the ratio at $p_T>$ 2 
GeV/$c$. A fit to a constant gives a value of the ratio equal to 
$0.81~\pm~0.02^{\rm stat}~\pm~0.09^{\rm syst}$ consistent with our 
previous measurement of 
$0.85~\pm~0.05^{\rm stat}~\pm~0.09^{\rm syst}$~\cite{ppg064}.  The 
{\sc pythia} prediction of the $\omega$/$\pi$ ratio, shown in 
Fig.~\ref{fig:omegapi0ratio} with a solid line, lies above the 
measured ratio.

The $\omega/\pi$ ratios measured in minimum bias $d+$Au, Cu$+$Cu, and 
Au$+$Au collisions at $\sqrt{s_{NN}}$=200 GeV are presented in 
Fig.~\ref{fig:omegapi0ratioA}. As in the case of $p+p$ collisions 
there is no indication that the ratios depend on transverse momentum 
for $p_T>$ 2 GeV/$c$. Fits to a constant for $p_T>$ 2 GeV/$c$ give the 
following values of the $\omega/\pi$ ratio:
$0.75~\pm~0.01^{\rm stat}~\pm~0.08^{\rm syst}$ in $d+$Au, 
$0.71~\pm~0.07^{\rm stat}~\pm~0.07^{\rm syst}$ in Cu$+$Cu and 
$0.83~\pm~0.09^{\rm stat}~\pm~0.06^{\rm syst}$ in MB Au$+$Au 
collisions. Within the uncertainties the $\omega/\pi$ ratios measured 
in different collision systems for $p_T>$ 2 GeV/$c$ are in agreement. 
This agrees with previous measurements in $d+$Au \cite{ppg064} within 
the uncertainties. The ratios in various collision systems imply 
similar suppression factors and $p_T$ dependences within the 
uncertainties for the $\omega$ and $\pi$ production in nucleus-nucleus 
collisions at high $p_T$.

\begin{figure}[th]
\includegraphics[width=1.0\linewidth]{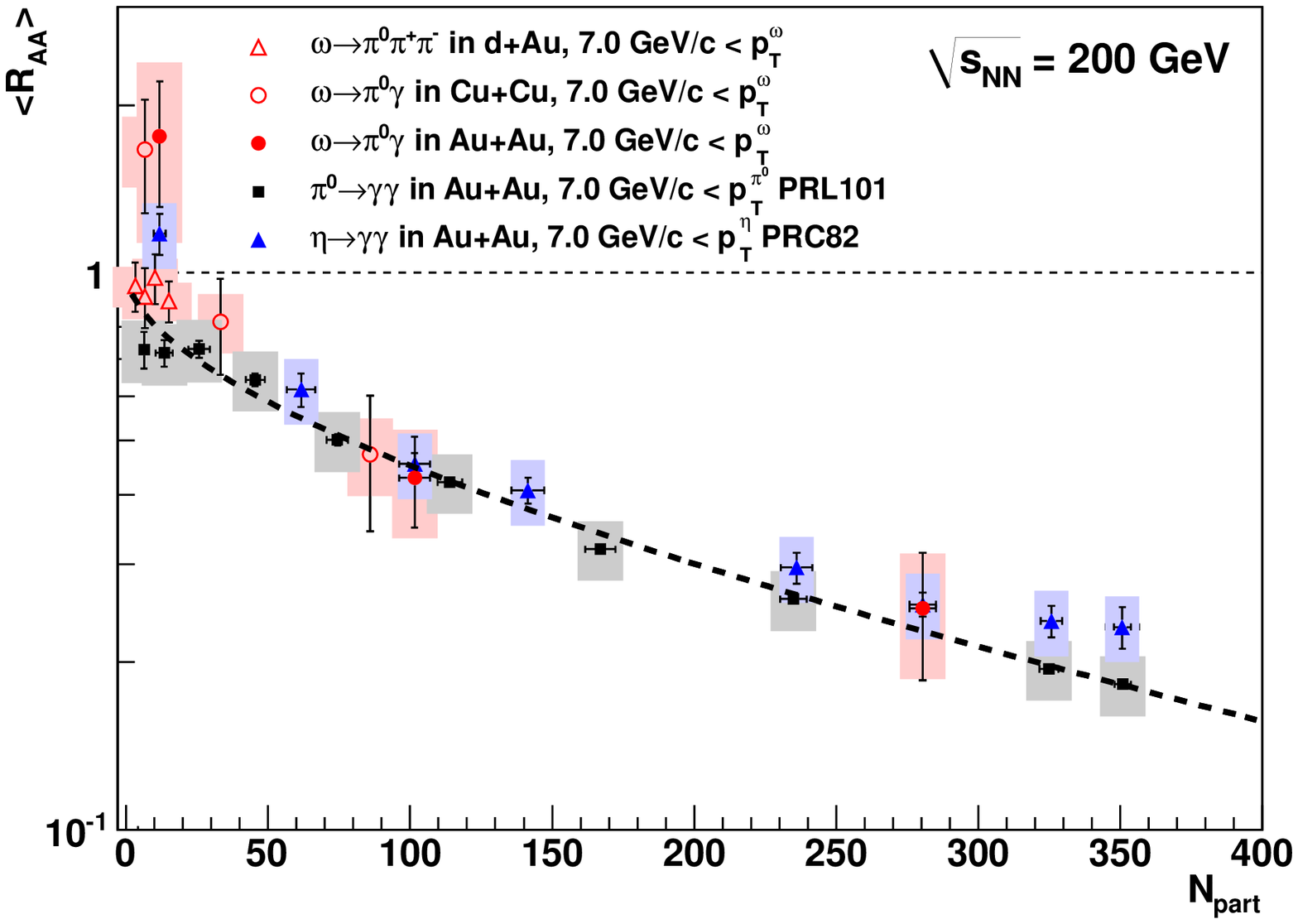}
\caption{\label{fig:Npart}
$R_{\rm AA}$ for the $\omega$ meson integrated over the range $p_T>$ 7 
GeV/$c$ as a function of the number participating nucleons 
($N_{\rm part}$). Results for $\pi^{0}$~(PRL101)~\cite{ppg080} and 
$\eta$~(PRC82)~\cite{ppg115} are shown for comparison. The dashed line 
shows the fitted fractional energy loss function, 
$R_{\rm AA} = (1-S_{0}N^{a}_{\rm part})^{n-2}$.
}
\end{figure}

\subsection{\label{subsec:raa}Nuclear modification factors}

To quantify medium induced effects on high $p_T$ particle production the 
nuclear modification factor is defined as:
\begin{equation}
R_{\rm AB}(p_T)=\frac{d^{2}N_{\rm AB}/dydp_T}{(\langle N_{\rm coll}\rangle 
/\sigma^{inel}_{pp})\times d^{2}\sigma_{pp}/dydp_T},
\end{equation}
where $d^{2}N_{\rm AB}/dydp_T$ is the differential yield per event in 
nucleus-nucleus collisions, $\langle N_{\rm coll} \rangle$ is the 
number of binary nucleon-nucleon collisions averaged over the impact 
parameter range of the corresponding centrality bin calculated by 
Glauber Monte-Carlo simulation \cite{Glauber}, and 
$\sigma^{inel}_{pp}$ and $d^{2}\sigma_{pp}/dydp_T$ are the total and 
differential cross sections for inelastic $p+p$ collisions, 
respectively. In the absence of medium-induced effects, the yield of 
high-$p_T$ particles is expected to scale with $\langle N_{\rm coll} 
\rangle$, resulting in $R_{\rm AB}$=1 at high-$p_T$.

Figure \ref{fig:omegaRdAu} presents $R_{d{\rm Au}}$ measured for the 
$\omega$ in minimum bias, most central and peripheral $d+$Au 
collisions at $\sqrt{s_{NN}}$=200 GeV. Good agreement is observed 
between different decay modes, and between new and previously 
published PHENIX $\omega$ results \cite{ppg064} shown with open 
markers. For comparison we also present $\pi^{0}$ results published in 
\cite{ppg044} in the figure. In peripheral collisions the measured 
values of $R_{d{\rm Au}}$ are consistent with unity over the whole 
$p_T$ range of measurements. In most central collisions a modest 
Cronin-like enhancement is observed in a range of $p_T$ from 2 GeV/$c$ 
to 6 GeV/$c$ and suppression of $\omega$ production at 

$p_T>8$~GeV/$c$. A similar enhancement at 2--6~GeV/$c$ was previously 
observed for neutral and charged pions~\cite{ppg030,ppg044} and $\phi$ 
mesons~\cite{ppg096}. Suppression of $\omega$ production at higher 
$p_T$ is in agreement with $\pi^{0}$ results \cite{ppg044}. Similarity 
of the observed effects for the mesons with very different masses 
suggests that the collective nuclear effects occur at the partonic 
level~\cite{recomodel1, recomodel2, Cronin:1974zm}.

Nuclear modification factors measured in Cu$+$Cu and Au$+$Au 
collisions at $\sqrt{s_{NN}}$=200 GeV as a function of $p_T$ are shown 
in Fig.~\ref{fig:omegaRAA}. Results are presented for minimum bias, 
most central (0--20\%), midcentral (20--60\%) and peripheral (60--94\% 
in Cu$+$Cu; 60--92\% in Au$+$Au) collisions. The nuclear modification 
factors do not depend on $p_T$ for $p_T>$ 6 GeV/$c$ at all 
centralities. For $N_{\rm part}>34$ suppression of $\omega$ production 
begins to be observed, with suppression increasing as $N_{\rm part}$ 
increases.

Figure \ref{fig:Npart} shows $R_{\rm AA}$ values integrated for $p_T>$ 
7 GeV/$c$ as a function the number of participants. For $\omega$ 
mesons we present four centrality bins in $d+$Au, and three centrality 
bins in Cu$+$Cu and Au$+$Au. For comparison the average values of 
$R_{\rm AA}$ for $\pi^{0}$ \cite{ppg080} and $\eta$ mesons 
\cite{ppg115} mesons for $p_T>$ 7 GeV/$c$ are also plotted. To see 
whether the $\omega$ follows the suppression pattern of $\pi^{0}$ and 
$\eta$, the integrated $R_{\rm AA}$ vs $N_{\rm part}$ dependence is 
fit to a fractional energy loss function $R_{\rm AA} = 
(1-S_{0}N^{a}_{\rm part})^{n-2}$ \cite{ppg080, whitepaper}. The 
parameter $n$, which is an exponent of the power law fit to the 
$\omega$ $p_T$ spectrum measured in $p+p$ for $p_T>$5 GeV/$c$ 
\cite{ppg099}, was fixed to 8. The fitting gives $\chi^{2}/$ndf less 
than three and parameters $S_{0}$ = (9.9$\pm$0.7)$\times$10$^{-3}$ and 
$a$ = 0.55 $\pm$ 0.01. As in \cite{ppg080} we find the parameter $a$ 
consistent with predictions of the GLV \cite{GLV} and PQM \cite{PQM} 
models ($a \sim 2/3$). Therefore, we can conclude that $\omega$ 
production has a similar suppression pattern as $\pi^{0}$ and $\eta$ 
which supports the scenario that the energy loss takes place at the 
parton level in the hot and dense medium formed in the collisions.

\section{\label{sec:sum}Summary}

We measured production of the $\omega$ meson via both leptonic and 
hadronic decay channels in $p+p$, $d+$Au, Cu$+$Cu and Au$+$Au at 
$\sqrt{s_{NN}}$=200 GeV. The invariant transverse momentum spectra 
show good agreement in different decay channels in $p+p$ and $d+$Au. 
The $R_{d{\rm Au}}$ shows a moderate Cronin like enhancement at 
intermediate $p_T$ 2--6 GeV/$c$ and suppression for $p_T>8$ GeV/$c$ in 
most central $d+$Au collisions. The measurement of the nuclear 
modification factor for the $\omega$ meson in Cu$+$Cu and Au$+$Au 
collisions show that $\omega$ production has a similar suppression 
pattern as the $\pi^{0}$ and $\eta$ within model agreement, thus 
supporting the scenario that the energy loss takes place at the 
partonic level in the hot and dense medium formed in the collisions.

\begin{acknowledgments}

We thank the staff of the Collider-Accelerator and Physics
Departments at Brookhaven National Laboratory and the staff of
the other PHENIX participating institutions for their vital
contributions.  We acknowledge support from the
Office of Nuclear Physics in the
Office of Science of the Department of Energy, the
National Science Foundation, Abilene Christian University
Research Council, Research Foundation of SUNY, and Dean of the
College of Arts and Sciences, Vanderbilt University (U.S.A),
Ministry of Education, Culture, Sports, Science, and Technology
and the Japan Society for the Promotion of Science (Japan),
Conselho Nacional de Desenvolvimento Cient\'{\i}fico e
Tecnol{\'o}gico and Funda\c c{\~a}o de Amparo {\`a} Pesquisa do
Estado de S{\~a}o Paulo (Brazil),
Natural Science Foundation of China (P.~R.~China),
Ministry of Education, Youth and Sports (Czech Republic),
Centre National de la Recherche Scientifique, Commissariat
{\`a} l'{\'E}nergie Atomique, and Institut National de Physique
Nucl{\'e}aire et de Physique des Particules (France),
Ministry of Industry, Science and Tekhnologies,
Bundesministerium f\"ur Bildung und Forschung, Deutscher
Akademischer Austausch Dienst, and Alexander von Humboldt Stiftung (Germany),
Hungarian National Science Fund, OTKA (Hungary),
Department of Atomic Energy and Department of Science and Technology (India),
Israel Science Foundation (Israel),
National Research Foundation and WCU program of the
Ministry Education Science and Technology (Korea),
Ministry of Education and Science, Russian Academy of Sciences,
Federal Agency of Atomic Energy (Russia),
VR and the Wallenberg Foundation (Sweden),
the U.S. Civilian Research and Development Foundation for the
Independent States of the Former Soviet Union,
the US-Hungarian Fulbright Foundation for Educational Exchange,
and the US-Israel Binational Science Foundation.

\end{acknowledgments}

\clearpage


\end{document}